OPEN

# Computer aided detection of tuberculosis on chest radiographs: An evaluation of the CAD4TB v6 system

Keelin Murphy[1*], Shifa Salman Habib[2], Syed Mohammad Asad Zaidi[2], Saira Khowaja[3,4], Aamir Khan[3,4], Jaime Melendez[5], Ernst T. Scholten[1], Farhan Amad[2], Steven Schalekamp[1], Maurits Verhagen[6], Rick H. H. M. Philipsen[5], Annet Meijers[5] & Bram van Ginneken[1]

**There is a growing interest in the automated analysis of chest X-Ray (CXR) as a sensitive and inexpensive means of screening susceptible populations for pulmonary tuberculosis. In this work we evaluate the latest version of CAD4TB, a commercial software platform designed for this purpose. Version 6 of CAD4TB was released in 2018 and is here tested on a fully independent dataset of 5565 CXR images with GeneXpert (Xpert) sputum test results available (854 Xpert positive subjects). A subset of 500 subjects (50% Xpert positive) was reviewed and annotated by 5 expert observers independently to obtain a radiological reference standard. The latest version of CAD4TB is found to outperform all previous versions in terms of area under receiver operating curve (ROC) with respect to both Xpert and radiological reference standards. Improvements with respect to Xpert are most apparent at high sensitivity levels with a specificity of 76% obtained at a fixed 90% sensitivity. When compared with the radiological reference standard, CAD4TB v6 also outperformed previous versions by a considerable margin and achieved 98% specificity at the 90% sensitivity setting. No substantial difference was found between the performance of CAD4TB v6 and any of the various expert observers against the Xpert reference standard. A cost and efficiency analysis on this dataset demonstrates that in a standard clinical situation, operating at 90% sensitivity, users of CAD4TB v6 can process 132 subjects per day at an average cost per screen of $5.95 per subject, while users of version 3 process only 85 subjects per day at a cost of $8.38 per subject. At all tested operating points version 6 is shown to be more efficient and cost effective than any other version.**

Tuberculosis (TB) remains one of the top ten causes of death worldwide with approximately 10 million cases in 2017, causing an estimated 1.6 million deaths[1]. Definitive diagnosis of TB is unfeasibly time-consuming, with the gold standard of sputum culture testing being expensive and requiring several weeks for a conclusive result. TB is known to be an extremely contagious disease and prompt diagnosis and treatment are required as a means of infection control. In recent years the Xpert MTB/RIF® (GeneXpert, Cepheid, Sunnyvale, CA, USA)[2], molecular test (Xpert) has become increasingly popular, with a high sensitivity and specificity[2], and endorsement from the World Health Organization (WHO) since 2010. However, with the majority of TB cases occurring in resource-constrained settings, the cost of the Xpert test remains comparatively high (from $13 in countries where concessional pricing is available[3–5], and $46-175 in private healthcare settings[6]). Obtaining sputum samples from large populations is additionally both time-consuming and logistically difficult while daily throughput is limited by the processing time of 2 hours for the Xpert test. The WHO has identified the need for a triage test to identify subjects for referral for a further confirmatory test such as Xpert[7]. The documentation from the WHO indicates that such a triage test should be accessible to users with a minimum of training with no requirement for sputum collection. It should be fast, robust, portable, easily maintained and inexpensive. Furthermore, it should have a minimum sensitivity of 90% when compared with the confirmatory test, and a corresponding minimum specificity of 70%. For these reasons, there is growing interest in the use of chest X-Ray (CXR) as a means of simple and efficient pre-screening of populations in advance of sputum testing[8,9]. While CXR is a sensitive tool for detection of pulmonary TB[9], the lack of medical expertise to interpret CXR images in low-resource, high-burden settings has limited its usage in the past. This has prompted the development of analytical software capable of identifying the presence of TB from CXR images. In this study we evaluate one such commercial software platform, CAD4TB v6, developed in association with Radboud University Medical Center, the Netherlands. The CAD4TB software is distributed by Delft Imaging Systems and is already in use in numerous settings worldwide where its performance has been previously studied[3,10–14]. In 2018 version 6 of the software was released, the first version to use deep-learning technology. This version of the software can interpret a CXR image in less than 15 seconds and is designed to work on subjects from age 4 years and upwards. In this work its performance relative to previous versions and expert human observers is studied.

[1]Radboud University Medical Center, 6525, GA, Nijmegen, The Netherlands. [2]Community Health Solutions, Karachi, 74000, Pakistan. [3]Interactive Research and Development, Karachi, 75190, Pakistan. [4]The Indus Health Network, Karachi, 75190, Pakistan. [5]Thirona, 6525 EC, Nijmegen, The Netherlands. [6]Universal Delft Ltd., Delft, Ghana. *email: Keelin.Murphy@radboudumc.nl

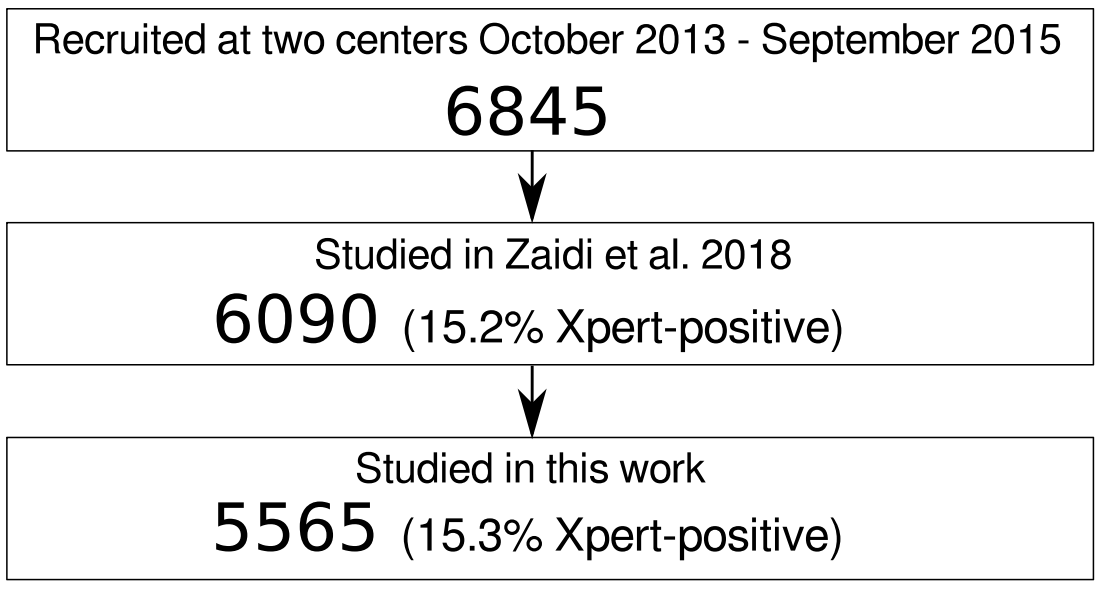

**Figure 1.** Flowchart describing data collection.

## Methods

**Data.** In this work CAD4TB is tested on data that is completely independent of the data used to train the system. The training process and data are not described for reasons of commercial sensitivity. The data used in this study were acquired from two purpose built TB treatment and diagnostic centers (known as Sehatmand Zindagi, "healthy life" centers) in Karachi, Pakistan between October 2013 and September 2015. Recruitment to these centers was via self-referral or through referral of individuals with presumptive TB (according to WHO screening recommendations[15]) from private health-care clinics in the locality. The reported symptoms of the participants were recorded including presence and duration of cough and presence of fever, haemoptysis or night sweats. Full analysis of symptoms can be obtained from previous work[14]. All participants underwent CXR and provided a sputum sample for Xpert MTB/RIF testing (hereafter referred to as Xpert) on the same day. Mucolator sachets were used to induce sputum production in cases where it could not be produced spontaneously. The CXR images were recorded digitally in dicom format. The results of the Xpert test are used as the reference standard throughout this work. Data from 6845 individuals was collected during the time period selected. The use of the data in previous[14] and current work is shown in Fig. 1. The frequency of Xpert-positive subjects is preserved in the dataset of 5565 subjects used in this study. This project was approved by the Institutional Review Board of Interactive Research and Development, Pakistan. Recruitment and experimental protocols were carried out in accordance with their guidelines and regulations. Informed consent was obtained from all participants or their legal guardians in the case of minors.

**Observer analysis.** To evaluate and compare the performance of human experts in detecting TB from this CXR dataset, 500 scans (250 Xpert positive and 250 Xpert negative) were selected for visual examination and scoring. To ensure that the selected set was not coincidentally more or less 'difficult' than average, it was chosen such that the performance of CAD4TB v6 on the 500 scans was equivalent to that on the entire set (Area under Receiver Operating Curve (ROC) = 0.885, as described in the Results section). This was done by repeated random sampling and testing until a set was found where the CAD4TB performance met the stated requirements.

The observers were shown each of the 500 scans on a browser-based platform where zooming, panning and window-levelling were permitted as desired. The observers were aware that the images came from a high-burden setting but blinded to clinical information and to each other's scores. Each observer assigned every scan one of the following scores:

1. No TB: Scan is normal or has an abnormality not related to TB
2. Possible TB: Scan has some abnormalities, TB cannot be ruled out
3. Likely TB: Scan has abnormalities which are strongly suggestive of TB

The observer could also add free text if desired, to indicate any other observations. Five observers with various backgrounds and experience, as described below, scored all 500 selected scans.

- Observer 1: A radiologist with special interest in chest radiology and more than 30 years of experience working in the Netherlands.
- Observer 2: A public health TB-doctor in the Netherlands with 25 years experience in chest X-ray reading.

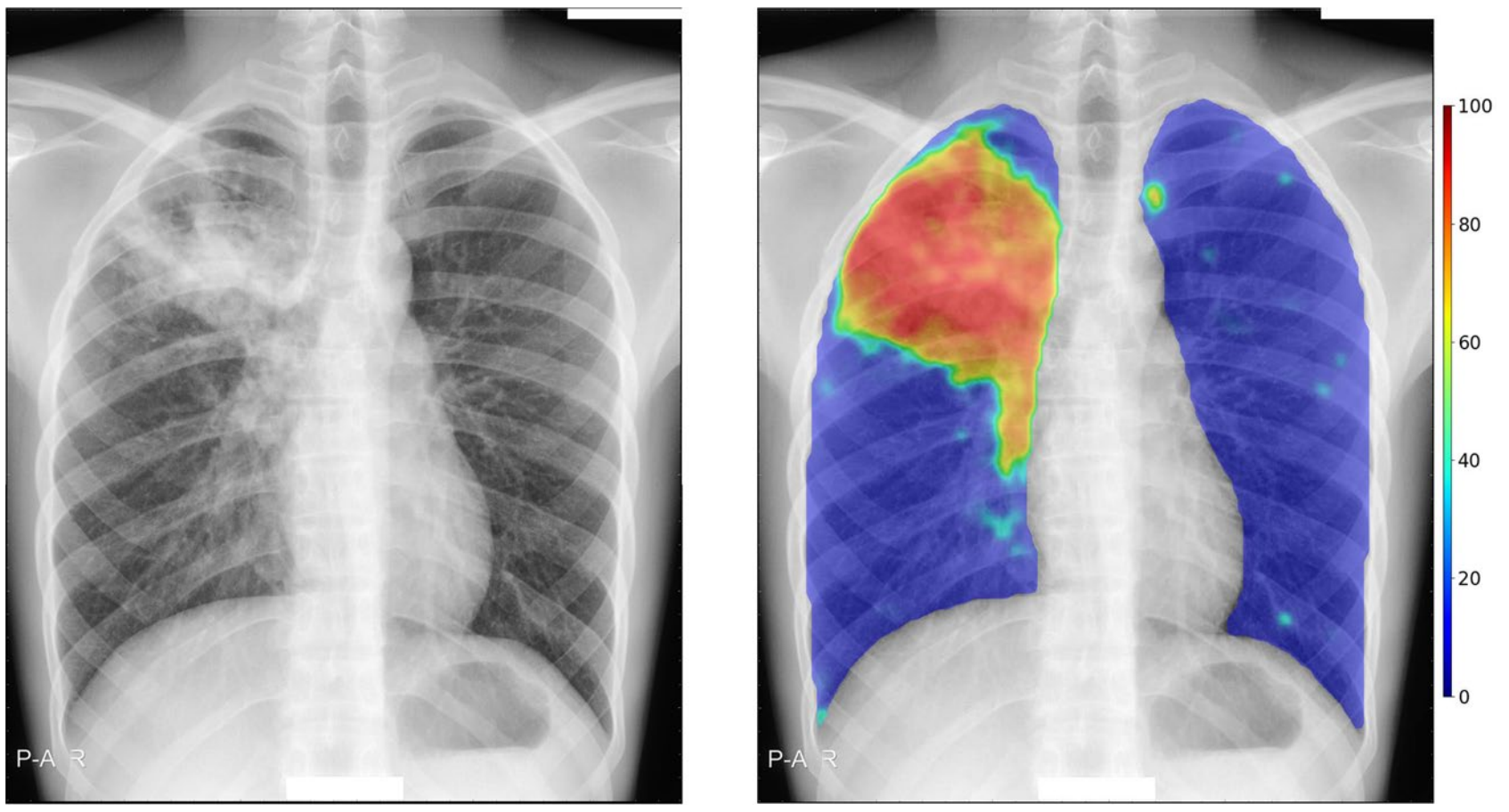

a b

**Figure 2.** Sample output from CAD4TB v6. (**a**) The original radiograph, (**b**) The radiograph with abnormality heatmap overlay. The final composite CAD4TB score for this subject was 91.7 (0 = normal, 100 = most abnormal) and the Xpert test was positive.

- Observer 3: A radiologist working as a consultant in chest radiology for a network of private TB diagnostic and treatment facilities in Pakistan
- Observer 4: A radiologist with specialty in chest radiology and 4 years of experience working in the Netherlands.
- Observer 5: An X-ray technician working at one of the 61 Sehatmand Zindagi health centers (static center) in Pakistan. Acquires more than 100 scans per day in this setting and operates the CAD4TB software.

**CAD4TB analysis.** CAD4TB is a commercial software platform which uses machine-learning techniques to automatically detect TB from CXR images. The software has been trained on independent annotated datasets to recognise distinctive features of TB in CXR images. It outputs a score (0–100), which may be interpreted as the probability that the subject is suffering from active TB visible on CXR. An abnormality heatmap indicating regions which the software considers suspicious is also produced, see Fig. 2. The most recent version of this platform, CAD4TB v6, released in 2018, uses deep-learning, a variant of machine-learning using deep neural networks. Deep learning has rapidly become the technique of choice in the field of computerized image interpretation and in the last decade has been repeatedly demonstrated to outperform other methods in a multitude of tasks and settings, including medical image analysis[16]. Four versions of CAD4TB are compared in this work (from oldest to newest: v3, v4, v5 and v6). Previous works have examined the performance of version 3[3,10,11,14] and version 5[12,13] in a variety of settings.

The processing pipeline for all versions begins in the same way with an energy-based normalization step[17] to reduce the difference in appearance between radiographs acquired from different machines. This is followed by a quality check to determine that the image to be processed is a valid postero-anterior chest radiograph. Next the lung fields are segmented to identify the region of interest. This step is improved by the use of a deep neural-network in version 6.

Following this initial pre-processing, analysis of the image abnormality is commenced. The principal analysis concerns the texture of the lung parenchyma which is handled by different types of classifier in various versions as follows: v3 - kNN classifier, v4 and v5 - support vector machine, v6 - deep convolutional neural network. In versions from 4 onwards a heatmap indicating regions of abnormality is produced. Versions 3 and 4 additionally use a symmetry check since asymmetry of the lung fields indicates abnormality[18]. All versions prior to version 6 also employ shape analysis, which was not found to be necessary in version 6 with the improved lung segmentation and texture analysis. The combination of all abnormality analyses is aggregated, in all versions (with the exception of version 6 which uses only texture analysis), to produce a final score (0–100) for likelihood that TB is present in the image.

In this work the software is run on all 5565 scans in the described dataset and the output scores are obtained and analyzed firstly with reference to the Xpert results and secondly with respect to a radiological reference standard set by expert observers. The radiological reference standard is set on the 500 scans described in the previous section which were read by 5 observers. The reference standard created identified a subject as TB positive if 3 or more of the 5 observers had scored the examination with scores 2 or 3. In this way 338 scans were marked positive and 162 negative. To compare the performance of CAD4TB v6 directly with each observer we further create 5 separate radiological reference standards as described above, but, in this case, each time we exclude a different observer. A scan is identified as TB positive if 2 or more of the remaining 4 observers gave a score of 2 or 3. Each resulting reference standard is then used to evaluate the performance of both CAD4TB v6 and the excluded observer. Confidence intervals in all cases are obtained by bootstrapping[19].

**Cost analysis.** As in previous work[3] we perform cost analysis based on a hypothetical point-of-care testing unit with one digital radiography system and three 4-cartridge GeneXpert IV machines. The digital radiography system is operated by a trained technician and all image analysis is done by CAD4TB. It has a capacity of 300 CXR screens per day while the Xpert testing capacity is 45 tests per day.

The overall cost for an Xpert test in the 145 countries eligible for subsidised Xpert testing has been estimated at $13.06 including equipment, resources, maintenance and consumables[3]. The sources for these prices remain unchanged[5] and the same figure is thus retained in this work. Similarly for the digital radiography system, the cost analysis used in previous work[3] is re-used, where equipment and running costs including labor, maintenance, consumables, depreciation etc. have been fully itemized resulting in a cost of $1.49 per CXR screening.

Since CXR screening is many times cheaper and faster than Xpert testing, we perform cost, efficiency and sensitivity analysis of a scenario where CAD4TB identifies a proportion of subjects eligible for subsequent Xpert testing and the remainder are discharged after the CXR screen. The CAD4TB system can be operated at varying levels of sensitivity (and associated specificity), thus we perform the same analysis at four different high-sensitivity settings, to illustrate the effect on cost and efficiency at each setting, for each version of CAD4TB. The following figures are calculated at each setting:

- $s$: System sensitivity. The proportion (0–1) of Xpert-positive cases that would be correctly identified by CAD4TB at this operating point
- $p_X$: The proportion (0–1) of cases that will be sent for subsequent Xpert testing . This is all cases that would be marked as TB positive by CAD4TB at this operating point (including some false positives).
- $C_{AVG}$: The average cost for a case arriving at the unit. $C_{AVG} = \$1.49 + (p_X \times \$13.06)$
- $C_{TB}$: The average cost per TB case detected. This calculation requires an estimate of TB prevalence, for which we use the incidence in our dataset. $C_{TB} = C_{AVG}/\frac{854}{5565}$
- $\theta$: The daily throughput, i.e. the number of subjects that can be screened per day. While several hundred CXR screens could be performed in a day, the actual throughput is limited by the capacity of 45 Xpert screens per day in all tested scenarios and is given by $\theta = 45/p_X$
- In contrast to the study data used by Philipsen *et al.*[3], we do not have a reference standard of sputum culture testing, therefore it should be noted that the sensitivity values calculated in this work are relative to the Xpert results only. The comparison is made, thus, between a setting where only Xpert testing is available and a setting where CAD4TB pre-screening is available in addition to Xpert.

## Results

**Data.** Data collection resulted in symptom details, chest x-ray and Xpert results for 5565 individuals, 854 of whom had positive Xpert results. The characteristics of the data collected are described in Table 1.

**CAD4TB performance.** The four versions of the CAD4TB software are compared in part (a) of Fig. 3 with the outcome of the Xpert test as a reference standard. The sensitivity and specificity of each version of CAD4TB is obtained at multiple operating points by applying various thresholds on the output score in order to produce an ROC curve for each system. The area under the ROC curve (Az) is also shown for each system as an overall measure of performance. It is clear that the latest version, CAD4TB v6 demonstrates a marked improvement compared to its predecessors with notably higher sensitivities possible without loss of specificity. In version 6, with sensitivity set at 90% the system can achieve 76% specificity. The performances of CAD4TB versions 4 and 5 are very similar to each other on this dataset. version 4 has a marginally larger area under the curve than version 5 and shows improved specificity at lower sensitivity levels compared to both versions 5 and 6. CAD4TB v3 has the poorest performance with reduced specificity at all sensitivity settings compared to its successors.

In part (b) of Fig. 3 the performances of the four versions of CAD4TB are compared with the radiological reference standard created from the 5 reader opinions. Considering each version individually, there is no overlap in the range of AUC values given by the 95% confidence intervals in Fig. 3(a) and those given in Fig. 3(b). The agreement with the radiological reference standard is, therefore, significantly higher than with the Xpert results. Again, CAD4TB v6 has a markedly improved performance compared to previous versions, with an Az value of 0.987, achieving 98% specificity at a set operating point of 90% sensitivity. The weakest performance is that of version 3 (Az = 0.896).

Figure 4 illustrates the performance of CAD4TB v6 and of each observer compared with the 'consensus' of the four other observers. In each case the scores of the observer not in the reference standard are thresholded at score values 1 and 2 to obtain two distinct operating points of sensitivity and specificity. These are plotted with the curve for CAD4TB v6 shown for comparison. CAD4TB v6 shows a statistically significant improvement in performance compared to Observer 5 (a non-physician) at high sensitivities (the observer operating point is below the 95% confidence interval for the system performance). Otherwise the performance of CAD4TB v6 is very similar to expert observers, particularly at high sensitivities, and no observer is seen to perform significantly better (above the 95% confidence interval) than CAD4TB v6 at any operating point.

**Expert observer analysis.** As described in the previous section, observer performance is analysed by plotting two distinct operating points for comparison with the ROC curve of CAD4TB. Performance of all observers and CAD4TB v6 is illustrated in Fig. 5 using Xpert testing as the reference standard. For optimal accuracy the CAD4TB curve is calculated using all 5665 cases, however we additionally show the curve using only the 500 cases selected for annotation. All observer scores are clustered in the region of the CAD4TB curve with predominantly overlapping inter-observer 95% confidence intervals. Observer 5, the X-ray technician without formal medical/radiology training, has the weakest performance with operating points below the CAD4TB curve. Observers 1, 2 and 4, all experts from a Western setting, have operating points very close to the curve. Observer

| | | XPERT MTB/RIF | |
|---|---|---|---|
| | All n(%) | Positive n(%) | Negative n(%) |
| | 5665 (100) | 854(15.3) | 4711(84.7) |
| **Gender** | | | |
| Male | 2756 (49.5) | 400 (14.5) | 2356 (85.5) |
| Female | 2809 (50.5) | 454 (16.2) | 2355 (83.8) |
| **Age (10–101)** | | | |
| <=25 | 1525 (27.4) | 356 (23.3) | 1169 (76.7) |
| 26-45 | 2079 (37.4) | 282 (13.6) | 1797 (86.4) |
| 46+ | 1961 (35.2) | 216 (11.0) | 1745 (89.0) |
| **Cough** | | | |
| None | 714 (12.8) | 47 (6.6) | 667 (93.4) |
| <2 Weeks | 4525 (81.3) | 745 (16.5) | 3780 (83.5) |
| >2 Weeks | 326 (5.9) | 62 (19.0) | 264 (81.0) |
| **Fever** | | | |
| Yes | 4213 (75.7) | 731 (17.4) | 3482 (82.6) |
| No | 1352 (24.3) | 123 (9.1) | 1229 (90.9) |
| **Haemoptysis** | | | |
| Yes | 739 (13.3) | 153 (20.7) | 586 (79.3) |
| No | 4826 (86.7) | 701 (14.5) | 4125 (85.5) |
| **Night Sweats** | | | |
| Yes | 1732 (31.1) | 339 (19.6) | 1393 (80.4) |
| No | 3833 (68.9) | 515 (13.4) | 3318 (86.6) |

**Table 1.** Characteristics of the data from 5565 individuals included in this study.

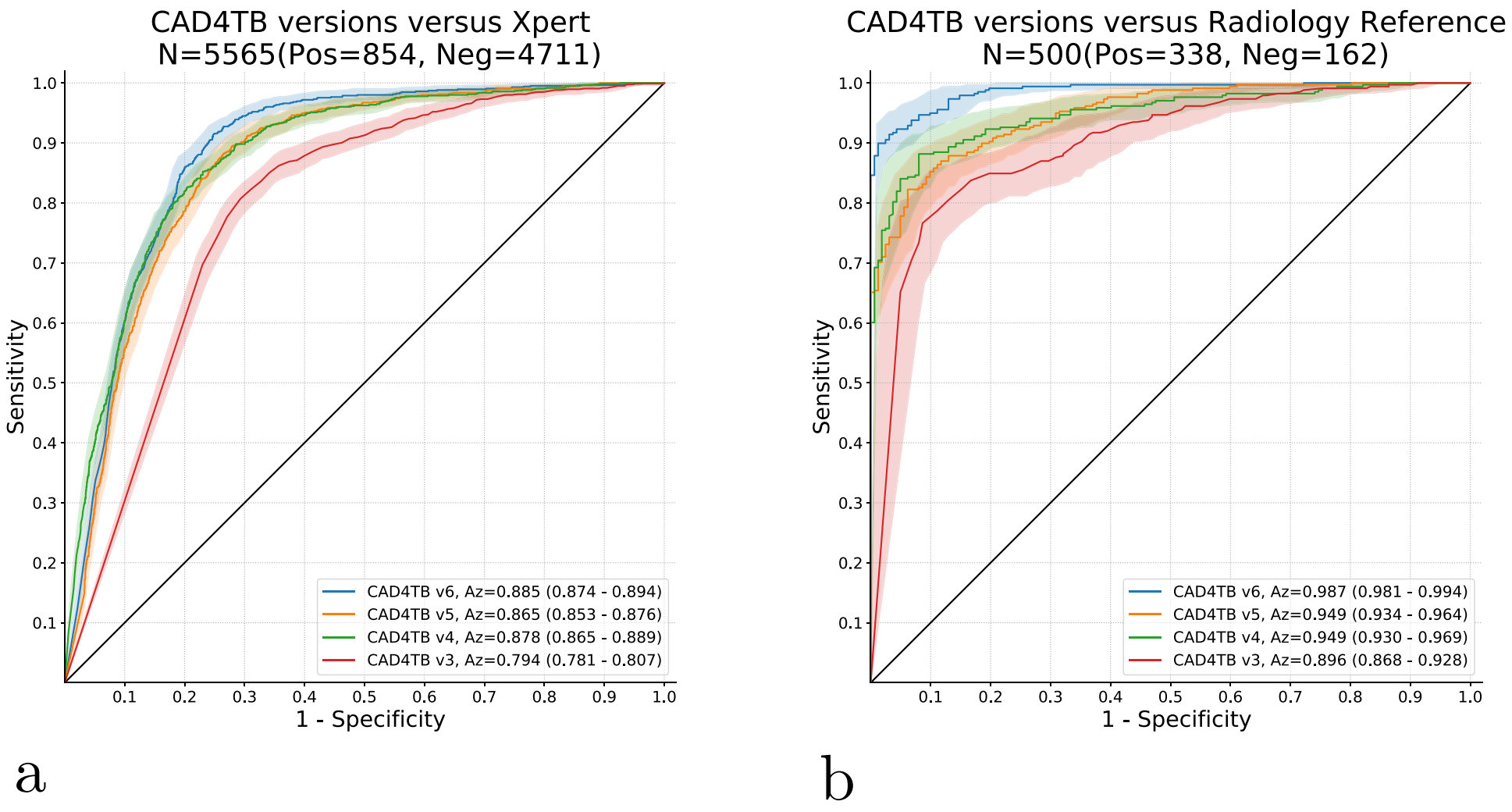

**Figure 3.** Comparing previous and current releases of CAD4TB. (**a**) Reference = Xpert. N = 5665. (**b**) Reference = Radiological 'Consensus'. N = 500. Shaded areas represent the 95% confidence intervals for the curves shown (calculated by bootstrapping).

1 retains a high sensitivity, close to 0.9, even at the higher threshold point (sensitivity = 0.88), implying a greater inclination to assign score 3 (strongly suggestive of TB). The radiologist experienced working in the studied high burden setting (Observer 3) has the best performance, particularly in terms of specificity, demonstrating notably improved specificity compared to all other observers at the lower threshold (where sensitivity varies very little - from 0.96 to 0.98).

**Cost analysis results.** The results of the analysis of cost, sensitivity and efficiency in the hypothetical point-of-care unit described in the Methods section are provided in Table 2. Statistical significance of cost differences with earlier versions (or with no CAD4TB triage) is tested by checking for overlap of ranges from 95% confidence intervals. CAD4TB v6 is significantly more cost-effective compared to version 3 in all cases, and at higher sensitivities it is also significantly improved compared to versions 4 and 5. CAD4TB v3 is the least cost-effective, while for this particular dataset the differences between versions 4 and 5 are marginal. Figure 6 graphically depicts the differences between versions 3 (the most commonly analysed version in the current literature) and version 6, at the four different levels of sensitivity. The figure also shows the contrast with the cost and throughput of the specified unit in the absence of CAD4TB screening.

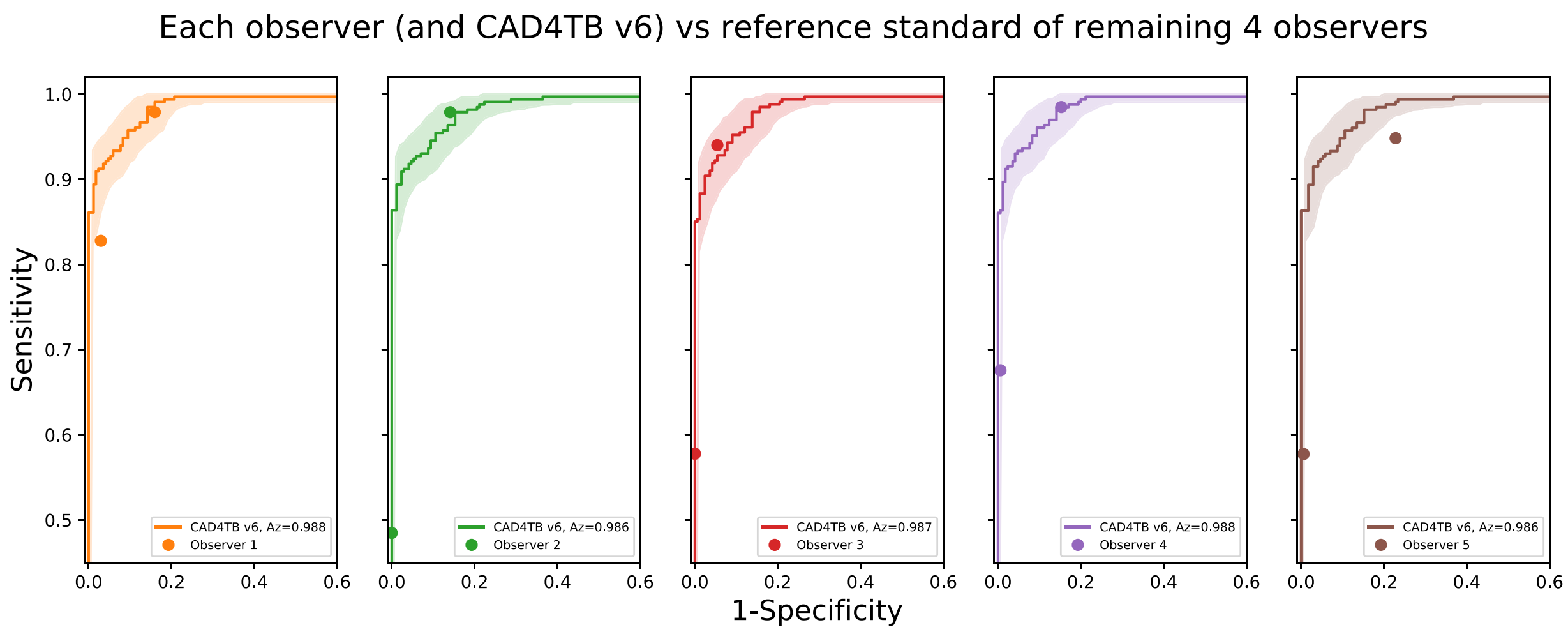

**Figure 4.** Each of the 5 observers is compared against a 'consensus' reference standard from the remaining 4 observers. In each case the scores of the observer not in the reference standard are thresholded at score values 1 and 2 to obtain two distinct operating points of sensitivity and specificity. The performance of CAD4TB v6 against the same reference standard is also illustrated in each case. Shaded areas representing 95% confidence intervals are calculated by bootstrapping.

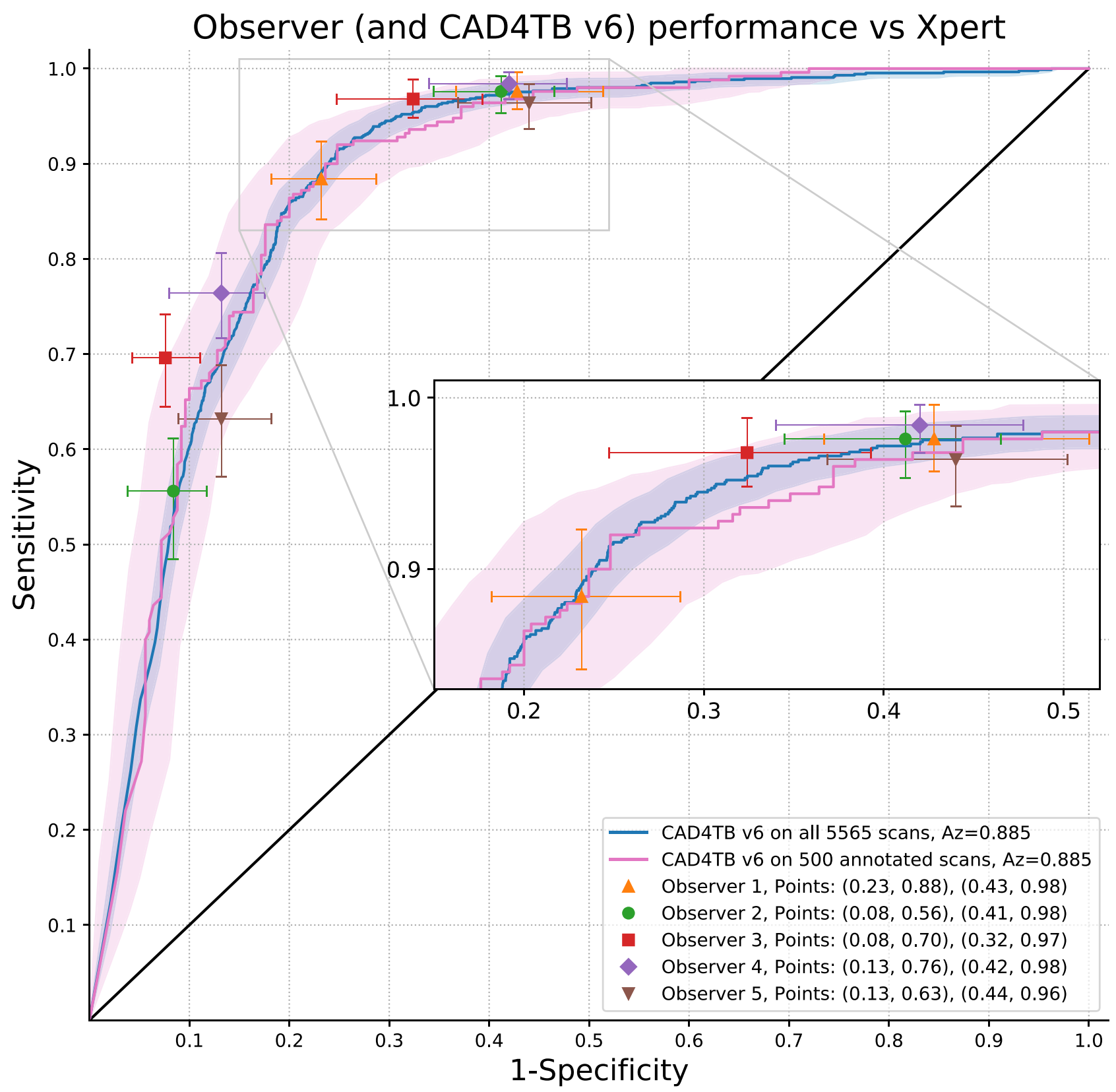

**Figure 5.** Expert observer performance compared with CAD4TB version 6, Reference = Xpert. The observers scored a set of 500 cases (250 positive, 250 negative). The CAD4TB curve is shown for both all 5665 cases (854 positive, 4711 negative), and the 500 annotated cases. (Note that the 500 cases were selected such that CAD4TB v6 Az = 0.885 over the set.) The shaded region of the curves and the error-bars on observer points represent 95% confidence-intervals calculated by bootstrapping.

| | **Without CAD4TB** | | | | |
|---|---|---|---|---|---|
| Sensitivity 1.0 | Specificity | 0.0 | | | |
| | $p_X$ | 100% | | | |
| | $C_{AVG}$ | $13.06 | | | |
| | $C_{TB}$ | $85.10 | | | |
| | $\theta$ | 45 | | | |
| | | CAD4TB v3 | CAD4TB v4 | CAD4TB v5 | CAD4TB v6 |
| Sensitivity 0.80 | Specificity | $0.712^{0}$ | $0.817^{0,3}$ | $0.793^{0,3}$ | **0.820**$^{0,3}$ |
| | $p_X$ | $36.7\%^{0}$ | $27.8\%^{0,3}$ | $29.8\%^{0,3}$ | **27.5%**$^{0,3}$ |
| | $C_{AVG}$ | $\$6.28^{0}$ | $\$5.12^{0,3}$ | $\$5.39^{0,3}$ | **$5.09**$^{0,3}$ |
| | $C_{TB}$ | $\$40.91^{0}$ | $\$33.35^{0,3}$ | $\$35.10^{0,3}$ | **$33.14**$^{0,3}$ |
| | $\theta$ | $123^{0}$ | $162^{0,3}$ | $151^{0,3}$ | **163**$^{0,3}$ |
| Sensitivity 0.85 | Specificity | $0.659^{0}$ | $0.770^{0,3}$ | $0.761^{0,3}$ | **0.804**$^{0,3}$ |
| | $p_X$ | $41.9\%^{0}$ | $32.5\%^{0,3}$ | $33.3\%^{0,3}$ | **29.7%**$^{0,3}$ |
| | $C_{AVG}$ | $\$6.96^{0}$ | $\$5.74^{0,3}$ | $\$5.84^{0,3}$ | **$5.36**$^{0,3}$ |
| | $C_{TB}$ | $\$45.37^{0}$ | $\$37.40^{0,3}$ | $\$38.05^{0,3}$ | **$34.96**$^{0,3}$ |
| | $\theta$ | $107^{0}$ | $138^{0,3}$ | $135^{0,3}$ | **152**$^{0,3}$ |
| Sensitivity 0.90 | Specificity | $0.540^{0}$ | $0.699^{0,3}$ | $0.706^{0,3}$ | **0.760**$^{0,3,4}$ |
| | $p_X$ | $52.8\%^{0}$ | $39.3\%^{0,3}$ | $38.7\%^{0,3}$ | **34.1%**$^{0,3,4}$ |
| | $C_{AVG}$ | $\$8.38^{0}$ | $\$6.62^{0,3}$ | $\$6.55^{0,3}$ | **$5.95**$^{0,3,4}$ |
| | $C_{TB}$ | $\$54.62^{0}$ | $\$43.14^{0,3}$ | $\$42.68^{0,3}$ | **$38.75**$^{0,3,4}$ |
| | $\theta$ | $85^{0}$ | $115^{0,3}$ | $116^{0,3}$ | **132**$^{0,3,4}$ |
| Sensitivity 0.95 | Specificity | $0.390^{0}$ | $0.591^{0,3}$ | $0.600^{0,3}$ | **0.688**$^{0,3,4,5}$ |
| | $p_X$ | $66.3\%^{0}$ | $49.2\%^{0,3}$ | $48.5\%^{0,3}$ | **41.0%**$^{0,3,4,5}$ |
| | $C_{AVG}$ | $\$10.14^{0}$ | $\$7.91^{0,3}$ | $\$7.82^{0,3}$ | **$6.84**$^{0,3,4,5}$ |
| | $C_{TB}$ | $\$66.09^{0}$ | $\$51.57^{0,3}$ | $\$50.95^{0,3}$ | **$44.56**$^{0,3,4,5}$ |
| | $\theta$ | $68^{0}$ | $91^{0,3}$ | $93^{0,3}$ | **110**$^{0,3,4,5}$ |

**Table 2.** Cost analysis. First section: without CAD4TB, second section: Using various versions of CAD4TB software as a triage tool. Four operating points with sensitivities from 0.80–0.95 are analysed. As per the text description $p_X$ = proportion of cases for subsequent Xpert testing, $C_{AVG}$ = Average cost per case, $C_{TB}$ = Average cost per TB case detected, $\theta$ = Daily throughput at the unit. Bold font represents the optimal value per row. Superscript $x$ indicates significant difference (based on 95% confidence interval range overlaps) compared to CAD4TB version $x$, and $x = 0$ indicates comparison with no CAD4TB triage.

## Discussion

In spite of many years of efforts to eradicate it, TB continues to be the leading cause of death from a single infectious agent worldwide. The WHO lists 30 high-burden countries which account for 87% of all cases worldwide. The availability of diagnostic solutions that are practical, affordable and efficient in these environments remains one of the greatest challenges to ending the global TB epidemic. In this work we demonstrated the efficacy of CAD4TB v6 as a pre-screening tool in high burden settings. Experiments have been carried out on a large and representative dataset, however we note that there are some limitations to the data. Firstly the utilized reference standard of Xpert test results is imperfect. The WHO estimate the pooled sensitivity and specificity values of Xpert for detection of TB as 92.5% and 98% respectively (based on 12 single centre evaluation studies)[5]. Nonetheless, in the absence of culture smears which are frequently too costly and time-consuming to obtain and analyse, it is common to rely on Xpert as a reference standard. In this work we benchmark against the WHO recommendations for minimum sensitivity and specificity values for a triage test as compared with a confirmatory test such as Xpert[7]. A second limitation is that the dataset comprises subjects that were symptomatic upon presentation to the clinics. This implies that TB prevalence in our study may be higher than in a random population from the same setting. In particular it should be observed that the cost per notified TB case may be increased in settings with lower prevalence. The dataset does not contain any information about the final diagnosis (if any) for TB-negative subjects which may be considered a limitation in terms of specificity analysis. There is also no recorded information regarding the HIV status of the subjects in this study, although Pakistan is a low-burden setting for HIV. It is well documented that TB is more difficult to detect on radiographs of HIV-positive subjects and therefore the CAD4TB results reported here may not generalize well to a high-burden HIV setting. Finally, we do not demonstrate any comparison with other machine-learning systems for detection of TB in chest radiographs and while our dataset is exceptionally large and has strong reference standards, both radiological and bacteriological, it must be noted that it represents a single population. For further information on the performance of CAD4TB v6 compared with other systems and on other datasets we refer the reader to a recent independent study from the Stop TB Partnership[20] which studies three TB-detection systems on 1196 subjects from two separate populations.

Version 6 of CAD4TB, analysed with Xpert as a reference standard, shows substantial improvement compared to previous versions in particular at higher sensitivity levels. This achievement is particularly important in the triage setting where the CAD4TB score is used as a selection criterion for subsequent sputum testing and/or treatment. A WHO consensus meeting to identify targets for new TB diagnostic tools in 2014 recommended that triage tests should have a sensitivity of 90% with a specificity of 70%[7] when compared with a confirmatory test such as Xpert. At the operating point with 90% sensitivity a specificity of 76% is achieved by CAD4TB v6 with all other versions having specificities at or below 70%. In our population of 5665 individuals this gain in specificity would represent a total saving of 283 Xpert tests.

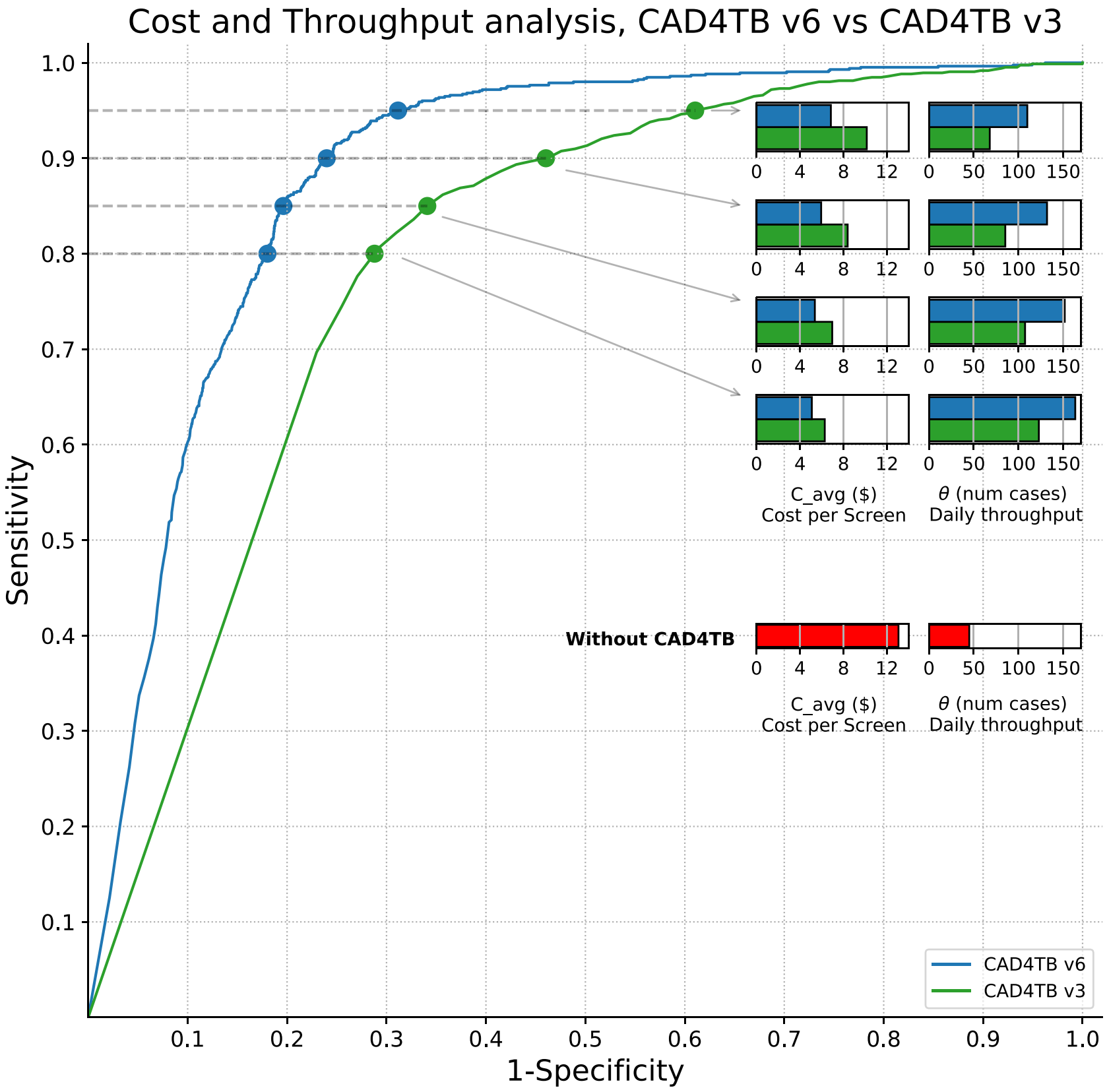

**Figure 6.** An illustration of how CAD4TB, used as a pre-screening tool, reduces costs and increases daily throughput. Results for both CAD4TB v6 and CAD4TB v3 are shown at four different sensitivity levels. The inset bar charts illustrate the average cost per screening ($C_{AVG}$) and the daily throughput of the unit ($\theta$) for each sensitivity level and CAD4TB version. Costing and throughput in the absence of CAD4TB is also illustrated.

While any triage test with a sensitivity below 100% will inevitably miss some cases of TB we note that if a triage test is easy to use, inexpensive and commonplace then the number of TB cases detected is likely to rise overall compared to a scenario where no triage testing is available[7]. The recommendations of the WHO also indicate that a triage test should be robust and portable. While x-ray machines are common in hospitals throughout the world there are many remote and low-resource communities where access to a hospital or medical center is impossible. CAD4TB is already in use around the world in mobile units including x-ray machines which are capable of operating without access to an electrical power grid. While these units can travel, even on unpaved roads, there are also x-ray machines available which are portable enough to be carried on foot and operate on battery, making x-ray a technology which could be made available to even the most remote of locations.

When compared with a radiological reference standard based on expert readings the performance of all CAD4TB versions is substantially improved compared to their performances against Xpert (Fig. 3). The fact that performance improves when the reference standard is radiological, and is indistinguishable from human expert performance implies that a large portion of CAD4TB errors in predicting the Xpert outcome arise from sources which cannot be eliminated by improving interpretation of the radiograph. These may include erroneous Xpert results or radiographs which are difficult to interpret for both radiologists and CAD4TB alike. In such cases the presence of TB may be difficult or impossible to visualize on the radiograph or the presence of old (inactive) TB or a different pathology may be indistinguishable from active TB on the image. Figure 7 illustrates some cases where the appearance of the radiograph is contrary to the result from the Xpert test - firstly a radiograph with an appearance consistent with TB, but where the Xpert test was negative, and secondly a radiograph which appears normal but the subject had a positive Xpert test. In both cases CAD4TB performs as expected and in agreement with expert observers. The reasons for the contrast between radiograph appearance and Xpert results for these particular cases are unknown. In contrast Fig. 8 shows two straightforward cases where the radiograph (interpreted by CAD4TB and by experts) and Xpert results were consistent.

CAD4TB is intended for use primarily in high-burden settings where radiological expertise is rarely available. For optimal results it is desirable that the system performs as well as a human observer in recognizing TB. In

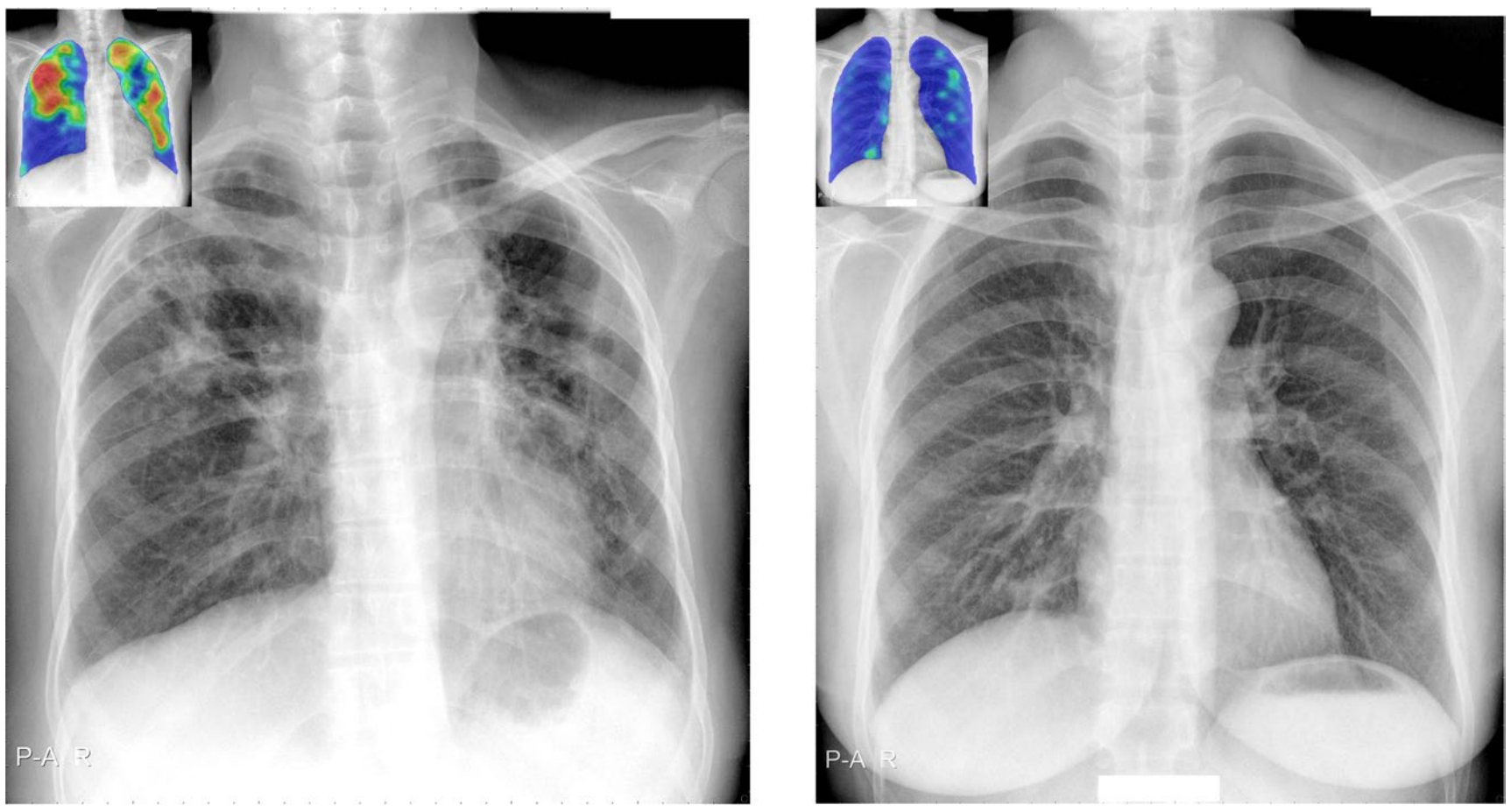

a b

**Figure 7.** Cases where radiograph presentation does not conform with Xpert result, making prediction by radiograph alone difficult for both observers and CAD4TB. The inset images show the CAD4TB heatmaps where blue indicates most normal texture and red indicates most abnormal. (**a**) An Xpert-negative case marked as TB positive (score 3) by all five observers and by CAD4TB v6 (score = 100). (**b**) An Xpert-positive case marked with score 1 (no-TB) by 4 of the experts and score 2 by the last one. The CAD4TB score for this case is 18.7 which is not picked up as TB positive until a sensitivity of 99% is reached.

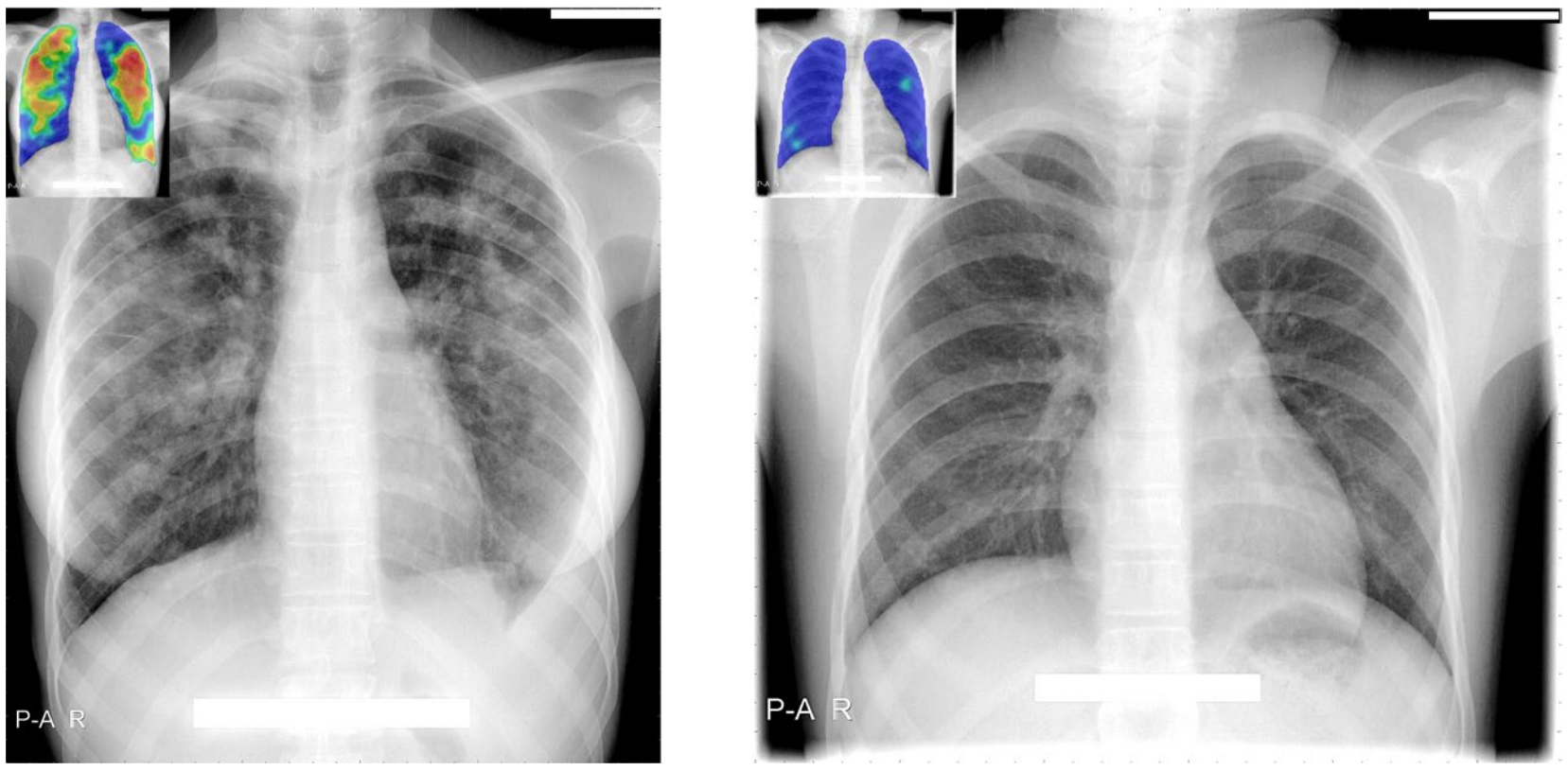

a b

**Figure 8.** Cases where radiograph interpretation by observers and CAD4TB conforms well with Xpert outcome. The inset images show the CAD4TB heatmaps where blue indicates most normal texture and red indicates most abnormal. (**a**) An Xpert-positive case marked as TB positive (score 3) by all five observers and by CAD4TB v6 (score = 91.7). (**b**) An Xpert-negative case marked with score 1 (no-TB) by all 5 experts. The CAD4TB score for this case is 7.1.

this work we compare the performance of CAD4TB v6 with that of five independent readers with various levels of expertise and experience in diagnosing TB on chest X-ray. In Figs. 4 and 5 it is clear that CAD4TB has a performance well within the range of the expert observers using both Xpert and radiological reference standards. Notably, the system improves on the performance of observer 5 in both Figs. 4 and 5. This observer is not trained as a physician or a radiologist and represents the typical skill level that might be expected from trained system operators in high-burden settings. From these results it is clear that at the most relevant, high, sensitivity levels the performance of CAD4TB version 6 is similar to that of human experts.

The cost and efficiency analysis in Table 2 demonstrates conclusively the advantage of CAD4TB version 6 over previous versions, and particularly if operating at higher sensitivities. At a sensitivity of 90% users of CAD4TB version 6 have a throughput that is 1.6 times higher, and a cost per screen that is 1.4 times lower than users of version 3. Figure 6 illustrates that at a very high sensitivity of 95% the cost per screened subject ($6.84) with CAD4TB v6 is almost half the cost at a unit without CAD4TB screening ($13.06) while the daily throughput of the unit (110) is almost 2.5 times higher.

In conclusion this work demonstrates that CAD4TB v6 is an accurate system, operating at the level of expert human readers in detecting TB from chest X-Ray. Used as a pre-screening system in regions where TB is endemic, CAD4TB allows for testing of much larger numbers of subjects at a fraction of the cost.

## Acknowledgements

This work was partially funded by Delft Imaging Systems. The funding body had no role in the decision to publish this study.

## Author contributions

K.M. analysed system performance and prepared manuscript including most figures S.S.H., S.M.A.Z., S.K., A.K. provided images and associated data, reviewed and edited manuscript J.M., R.H.H.M.P., A.M. Supplied output from CAD4TB system, assisted with figure preparation, reviewed and edited manuscript E.T.S., F.A., S.S., M.V. Participated as expert observers analysing 500 scans, reviewed and edited manuscript B.vG Supervised content and development of manuscript, reviewed and edited same.

## Competing interests

J.M., R.H.H.M.P., A.M. were in the employment of Thirona (developer of CAD4TB software) at the time of manuscript preparation. B.vG receives royalties and funding from Delft Imaging Systems and Mevis Medical Solutions and stock, royalties and funding from Thirona. The other authors report no conflicts.

## Additional information

**Correspondence** and requests for materials should be addressed to K.M.